\documentclass{aa}
\usepackage{graphics}
\def\a{{$\alpha$}}

\def\gsnr{{G~17.4--2.3}}
\newcommand{\h}{$^{\rm h}$}
\newcommand{\m}{$^{\rm m}$}
\newcommand{\s}{$^{\rm s}$}
\newcommand{\dd}{$\delta$}
\newcommand{\ha}{\rm H$\alpha$}
\newcommand{\hbeta}{\rm H$\beta$}
\newcommand{\HII}{\ion{H}{ii}}
\newcommand{\hnii}{{\rm H}$\alpha+[$\ion{N}{ii}$]$}
\newcommand{\nii}{$[$\ion{N}{ii}$]$}
\newcommand{\sii}{$[$\ion{S}{ii}$]$}
\newcommand{\oi}{$[$\ion{O}{i}$]$}
\newcommand{\oii}{$[$\ion{O}{ii}$]$}
\newcommand{\oiii}{$[$\ion{O}{iii}$]$}
\newcommand{\snr}{\rm supernova remnant}

\newcommand{\et}{et al.}
\newcommand{\flux}{$10^{-17}$ erg s$^{-1}$ cm$^{-2}$ arcsec$^{-2}$}
\newcommand{\dens}{\rm cm$^{-3}$}
\newcommand{\vel}{\rm km s$^{-1}$}
\newcommand{\sulfur}{[S~{\sc ii}]}

\newcommand{\siirat}{$[$\ion{S}{ii}$]\lambda\ 6716/6731$} 
\begin{document}

%
\title{First optical light from the supernova remnant G 17.4$-$2.3}

\author{P. Boumis\inst{1}
\and  F. Mavromatakis\inst{1}
\and E. V. Paleologou\inst{2}
}
\offprints{P. Boumis,~~\email{ptb@physics.uoc.gr}}
\authorrunning{P. Boumis et al.}
\titlerunning{First optical light from the supernova remnant G 17.4$-$2.3}
\institute{
University of Crete, Physics Department, P.O. Box 2208, 710 03 Heraklion, Crete, Greece 
\and Foundation for Research and Technology-Hellas, P.O. Box 1527, 711 10 Heraklion, 
Crete, Greece}
\date{Received 16 November 2001 / Accepted 29 January 2002}

\abstract{Deep optical CCD images of the supernova remnant \gsnr~were
obtained and faint emission has been discovered. The images, taken in
the emission lines of
\hnii, \sulfur~and \oiii, reveal filamentary structures in the
east, south--east area, while diffuse emission in the south and
central regions of the remnant is also present. The radio emission in
the same area is found to be well correlated with the brightest
optical filament. The flux calibrated images suggest that the
optical filamentary emission originates from shock-heated gas
(\sulfur/\ha\ $>$ 0.4), while the diffuse emission seems to
originate from an \HII\ region (\sulfur/\ha\ $<$ 0.3). Furthermore,
deep long--slit spectra were taken at the bright \oiii\ filament and
clearly show that the emission originates from shock heated gas. The
\oiii\ flux suggests shock velocities into the interstellar
$``$clouds'' greater than 100 \vel, while the \siirat\ ratio indicates
electron densities $\sim$240 cm$^{-3}$. Finally, the \ha\ emission has
been measured to be between 7 to 20 $\times$ \flux.
}

\maketitle

\section{Introduction}
The Galactic supernova remnants (SNRs) have been identified by both
radio (non-thermal synchrotron emission) and optical (optical emission
lines) surveys. New searches in both wavebands continue to identify
galactic SNRs (Fesen \& Hurford 1995, Fesen et al. 1997; Green 2001
and references therein; Mavromatakis et al. 2000, 2001, 2002) but
since the last few years, observations in X-rays have also detected
new SNRs (e.g. Seward et al. 1995). The ratio of
\sulfur/\ha~has become the standard discriminator used in optical SNR
observations because the photoionized nebulae (like \HII~regions and
planetary nebulae) usually exhibit ratios of about 0.1-0.3, while
collisionally ionized nebulae (like known Galactic SNRs) show ratios
typically greater than 0.4 (Smith et al. 1993). Fesen et al. (1985)
suggested that a division at \sulfur/\ha~$<$ 0.5 does not provide
clear evidence to distinguish SNRs from photoionized regions and
additional observations of the strong forbidden oxygen lines (\oi,
\oii~and \oiii) are needed to give a complete diagnostic. Furthermore,
theoretical shock models , generally predicted \sulfur/\ha~ratios of
0.5 to 1.0 for SNRs (Raymond 1979, Shull \& McKee 1979).
\par
G 17.4$-$2.3 is not a well known SNR, and was first detected by
Reich et al. (1988) in their Effelsberg 2.7--GHz survey, while its
radio image was published by Reich et al. (1990). It is classified as a
circular supernova remnant with an incomplete radio shell,
characterized by diffuse shell--like emission, an angular size of
$\sim$24\arcmin\ and a radio spectral index of $\sim$0.8 (Green
2001). Case \& Bhattacharya (1998) calculated its surface brightness
to be $1.3 \times 10^{-21}$ W m$^{-2}$ Hz$^{-1}$ m$^{-1}$. Because, there
is no direct distance determination, they have made an estimation
by utilizing the radio surface brightness -- diameter relationship
($\Sigma-D$) and found a distance of 8.5 kpc, but still the
uncertainties are large ($\sim$40\%). Green et al. (1997), through their
survey with the Parkes 64 m telescope, detected maser OH (1720
MHz) emission.  In radio surveys of the surrounding region, no pulsar
was found to be associated with \gsnr~but another SNR has been discovered
in its neighborhood. G 17.8$-$2.6 has a very well defined shell, it
lies about 30\arcmin~north--east of \gsnr~and has an angular diameter
of 24\arcmin~(Reich et al. 1988). Neither of these remnants has been
detected optically in the past and from our observations no optical
emission has been found in G 17.8--2.6. On the other hand, X--ray
emission was not detected from \gsnr~in the ROSAT All--sky survey,
while there is some evidence of X--ray emission from the neighboring
SNR G 17.8--2.6.

In this paper, we report the discovery of faint optical filaments from
\gsnr. We present \hnii, \sii~and \oiii~images which show
filamentary structure along the south--east edge of the remnant
correlated very well with the radio emission. Spectrophotometric
observations of the brightest filament were also obtained and the
emission lines were measured. In Sect.2, we present
informations concerning the observations and data reduction, while the
results of our imaging and spectral observations are given in Sect. 3
and 4, respectively. In the last section (Sect. 5) we discuss the physical
properties of \gsnr. 

\section{Observations}
\subsection{Imagery}
The observations presented here were performed with the 0.3 m
Schmidt-Cassegrain (f/3.2) telescope at Skinakas Observatory in Crete,
Greece in August 20 and 21, 2001. The 1024$\times $1024
(with 19$\times$19 $\mu$m$^{2}$ pixels) Tektronix CCD camera was used 
resulting in a scale of 4\arcsec.1 pixel$^{-1}$~and a field
of view of 70\arcmin~$\times$~70\arcmin.

Two exposures in \oiii\ and \sii\ of 2400 s each were taken during
the observations, while one exposure of 1800 s and one of 2400 s were
obtained with the \hnii\ filter. Note that the final images in each
filter are the average of the individual frames.
\par
The image reduction (bias subtraction, flat-field correction) was
carried out using the standard IRAF and MIDAS and their negative
gray--scale representation using the STARLINK Kappa and Figaro
packages. The astrometry information was calculated for each image
individually using stars from the Hubble Space Telescope (HST) Guide
Star Catalogue (Lasker et al. 1999). The spectrophotometric standard
stars HR5501, HR7596, HR7950, and HR8634 (Hamuy et al. 1992;
1994) were used for absolute flux calibration. All coordinates quoted
in this paper refer to epoch 2000.
\subsection{Spectroscopy}
Low dispersion long--slit spectra were obtained with the 1.3 m
telescope at Skinakas Observatory in 2001 August 21. The 1300 line
mm$^{-1}$~grating was used in conjunction with a 2000$\times$800 SITe
CCD (15$\times$15 $\mu$m$^{2}$~pixels) which resulted in a scale of
1 \AA\ pixel$^{-1}$ and covers the range of 4750 \AA\ -- 6815 \AA.  The
slit width is 7\farcs7 and it was oriented in the south--north
direction, while the slit length is 7\farcm9. The coordinates of
the slit centre are $\alpha =$ 18\h31\m30\s\ and $\delta =$
$-$14\degr51\arcmin33\arcsec\ and two spectra of 3600 s each were
obtained. The spectrophotometric standard stars HR5501, HR7596,
HR9087, HR718, and HR7950 were observed in order to calibrate the
spectra of \gsnr.
  \begin{table}
      \caption[]{Typically measured fluxes in the area of the brightest filament}
         \label{fluxes}
\begin{flushleft}
\begin{tabular}{lllll}
            \hline
            \noalign{\smallskip}
     & N$^{\rm 1}$  & SW$^{\rm1}$ & Center$^{\rm 1}$ & \HII~$^{\rm 2}$\cr	
\hline
\hnii\ 	& 116.0	& 111.2 & 108.0 & 99.7 \cr	
\hline
\sii\ 	& 16.3 & 22.1 & 17.3 & 8.6 \cr	
\hline
\oiii\ 	& 13.0	& 4.7 & 22.1 &  -- \cr	
\hline
\end{tabular}
\end{flushleft}
${\rm }$ Fluxes in units of \flux \\\
$^{\rm 1}$Median values over a 36\arcsec $\times$ 21\arcsec\ box \\\
$^{\rm 2}$Median values over a 60\arcsec $\times$ 60\arcsec\ box \\\
 \end{table}
%
%
  \begin {figure}
   \resizebox{\hsize}{!}{\includegraphics{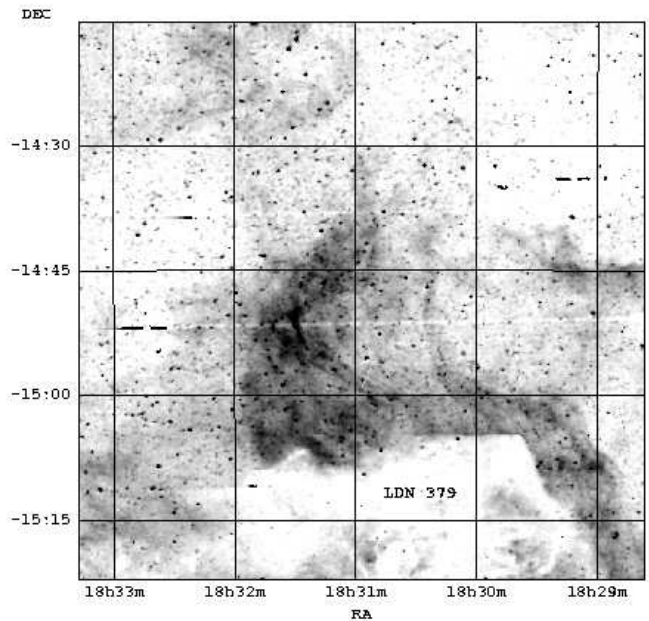}}
    \caption{ The field of \gsnr\ in the \hnii\ filter. 
     The image has been smoothed to suppress the residuals from the
     imperfect continuum subtraction.  Shadings run linearly from 0 to
     120$\times$ \flux.  The line segments seen near overexposed stars
     in this figure and the next figures are due to the blooming
     effect. 
      } 
     \label{fig01}
  \end{figure}
  \begin {figure}
   \resizebox{\hsize}{!}{\includegraphics{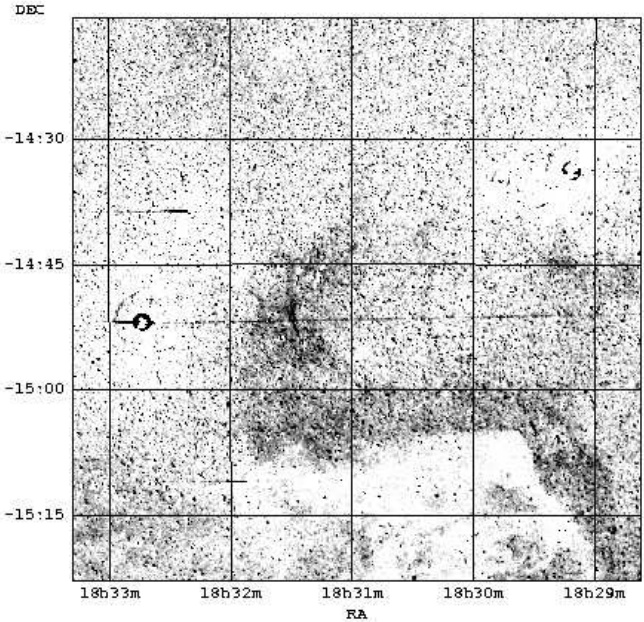}}
    \caption{ The \sii\ image of the area around \gsnr, 
     which has been smoothed to suppress the residuals 
     from the imperfect continuum subtraction. 
     Shadings run linearly from 0 to 20$\times$ \flux. 
     } 
     \label{fig02}
  \end{figure}
  \begin {figure}
   \resizebox{\hsize}{!}{\includegraphics{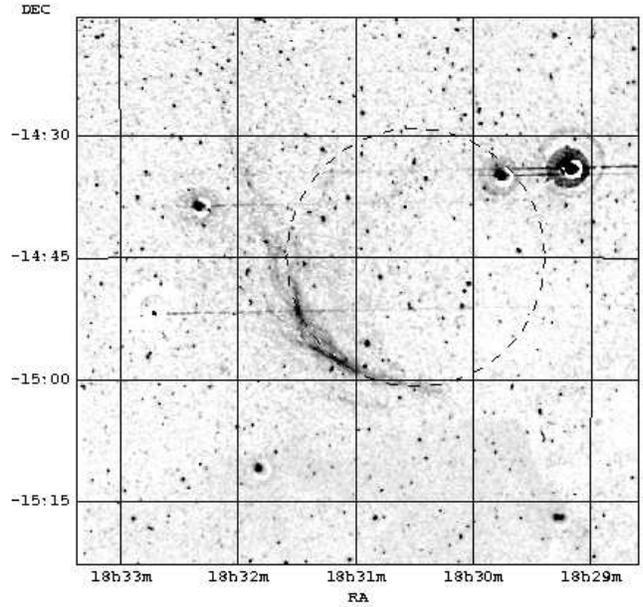}}
   \caption{ \gsnr\ imaged with the medium ionization line of \oiii\
   5007\AA. The image has been smoothed to suppress the residuals from
   the imperfect continuum subtraction and the shadings run linearly
   from 0 to 30$\times$ \flux. The projection of a
   $\sim$26\arcmin\ diameter sphere that matches the location and
   orientation of the filament is also shown.
     } 
     \label{fig03}
  \end{figure}
\section{Imaging of \gsnr}
\subsection{The \hnii\ and \sii\ emission line images}
The major characteristic revealed from our \hnii\ and \sulfur\ images
(Fig. \ref{fig01} and \ref{fig02}, respectively) seems to be the low
surface brightness of \gsnr.  The weak diffuse emission is present in
the south, south--east and central areas of the remnant, while no
emission is detected in the north--west part. The most interesting
region lies in the east part of the remnant, where a filamentary
structure exists, which is very well correlated with the radio
emission. In Table
\ref{fluxes}, we present typical average fluxes measured in several
locations within the field of \gsnr\ including the unknown
\HII~region which is located south--east of the remnant. A deeper study
of these images shows that the emission from the brightest part of the
remnant (east filament) originates from shock heated gas since we
estimate a ratio \sulfur/\ha\ $\sim$0.4--0.6, while a photoionization
mechanism produces the south--east \HII\ region (\sii/\ha\
$\sim$0.2--0.3). The possibility that the \HII\ emission contaminates
the remnant's emission can not be ruled out since for some of the
areas close to the filament as well as to the central region of the
remnant, we estimate \sulfur/\ha\ $\sim$0.2--0.3. The known dark,
extended nebula LDN 379 (Lynds \cite{lyn65}), which is at a distance
of $\sim$200 pc (Hilton et al. \cite{hil86}) is also visible in the
low ionization images.
\subsection{The \oiii\ 5007\AA~image}
\begin {figure}
   \resizebox{\hsize}{!}{\includegraphics{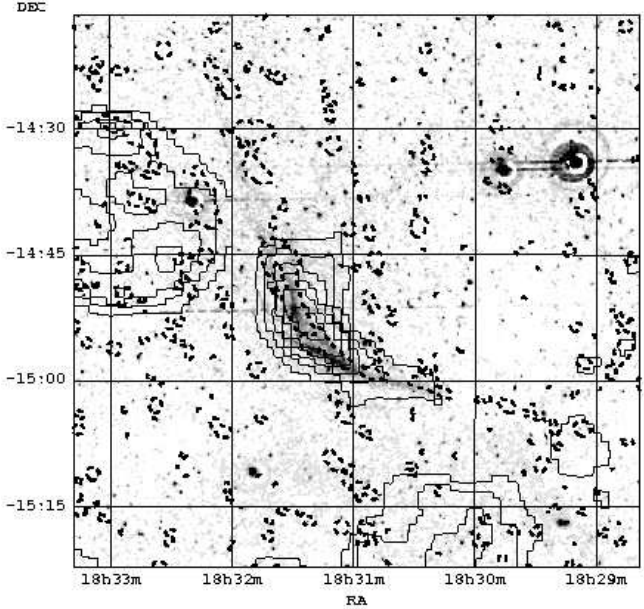}}
    \caption{The correlation between the \oiii\ emission and the radio
    emission at 1400 MHz (dash line) and 4850 MHz (solid line) is
    shown in this figure. The 1400 MHz (Condon \et\ \cite{con98}) and
    the 4850 MHz (Condon \et\ \cite{con94}) radio contours scale
    linearly from 1.1$\times$10$^{-3}$ Jy/beam to 0.1 Jy/beam and
    3.5$\times$10$^{-2}$ Jy/beam to 0.3 Jy/beam, respectively. LDN 379
    is also a strong radio source although not detected in \oiii. Note
    that in the north--east area of the filament we find the well--defined
    radio shell of the SNR G 17.8$-$2.6, which is also not detected in the
    optical.}
     \label{fig04}
  \end{figure}

In contrast to the previous results, the image of the medium
ionization \oiii\ line shows clearly filamentary nature of the
observed emission (Fig. \ref{fig03}). This bright filament extends for
$\sim$24\arcmin\ in the east, south--east, while no significant
emission was found in other areas of the remnant. We do not detect
\oiii\ emission where the diffuse \hnii\ and \sii\ emission is
detected. Table
\ref{fluxes}~lists also typical \oiii\ fluxes measured in different parts
of the filament. The latter, matches very well with the radio maps of
\gsnr\ at 1400 MHz and 4850 MHz, suggesting their association 
(Fig.~\ref{fig04}). The correlation shows that the filament is located
close to the outer edge of the radio contours but the low resolution
of the radio images does not help us to determine the actual position
of the filament with respect to the shock front. The geometry of the
\oiii\ filament allows us to define a $\sim$26\arcmin\ circle in diameter
(dashed circle in Fig. \ref{fig03}) for the remnant, with its center
at \a\ $\simeq$ 18\h30\m23.3\s, \dd\ $\simeq$
$-$14\degr45\arcmin25\arcsec. Note that the optical angular size is in
very good agreement with the value of $\sim$24\arcmin\ given in
Green's catalogue (Green 2001). However, a larger angular
diameter cannot be excluded since X--ray emission has not been detected
so far and the radio shell is incomplete.

\section{The long--slit spectra from \gsnr}
\begin {figure}
   \resizebox{\hsize}{!}{\includegraphics{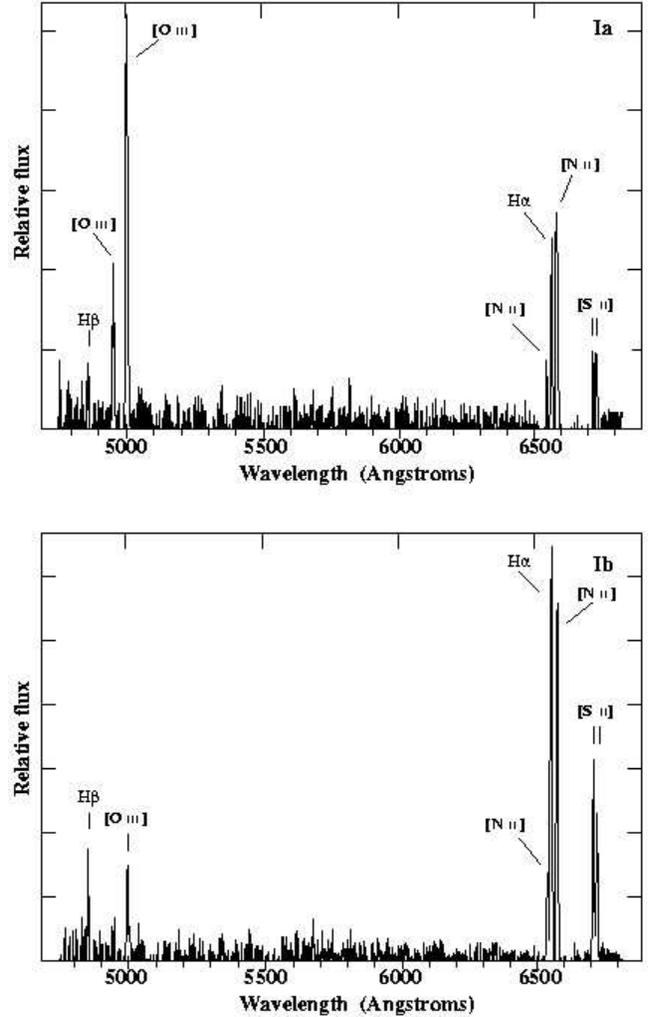}}
    \caption{The spectra of apertures Ia (top) and Ib (bottom).}
     \label{fig05}
  \end{figure}

The low resolution spectra were taken on the relatively bright optical
filament in the east part of the remnant (its exact position is given
in Sect. 2.2). In Table~\ref{sfluxes}, we present the relative line
fluxes taken from two different apertures (Ia and Ib,
Fig.~\ref{fig05}) along the slit. In particular, apertures Ia and Ib
have an offset of 9\arcsec.5 and 118\arcsec.5 north of the slit
centre, respectively. The apertures Ia and Ib were selected
because they are free of field stars in an otherwise crowded field. In
addition, these apertures include sufficient line emission, especially
in the blue part of the spectrum, to allow an accurate determination
of the observed lines. The background extraction aperture was taken
towards the north end of the slit. We utilized the flux calibrated
images to identify the nature of this background emission used in the
spectra. The \sii/\ha\ ratio of $\sim$0.2 measured in the images
suggests an \HII\ origin for the background aperture
emission. Assuming that its strength does not change appreciably at
the Ia and Ib positions we performed the background subtraction to
identify the SNR emission. The signal to noise ratios presented in
Table~\ref{sfluxes}~do not include calibration errors, which are less
than 10\%. Both extracted apertures show clearly that the observed
optical emission originates shock heated gas, since the \sii/\ha\ $>$
0.7. The \siirat\ ratio of 1.2--1.4 indicates low electron
densities (Osterbrock 1989).
\par
The very strong \oiii\ emission detected in aperture Ia suggests a
shock velocity greater than 100 \vel\ (Cox \& Raymond 1985), while the
sulfur lines ratio indicate an electron density $\sim$240
cm$^{-3}$~(Osterbrock 1989). However, taking into account the
statistical errors on the sulfur lines, electron densities up to 400
cm$^{-3}$~are compatible with our measurements. The shock velocity
implies by the Ib spectrum could be less but still around 100 \vel,
while the electron density is even lower than in Ia (less than 120
cm$^{-3}$).

\section{Discussion}
The \snr\ \gsnr\ shows up as an incomplete shell in the radio bound
without any X--ray emission detected so far. The low ionization images
generally show diffuse emission in the south and south--east areas of
the remnant. In contrast, a filamentary structure has been discovered
in the medium ionization \oiii\ line in the east, south--east region
which is very well correlated with the radio emission at 1400 and 4850
MHz and could define the remnant's outer edge. This correlation
indicates that the observed emission is associated to
\gsnr. Both the calibrated images and the long--slit spectra suggest
that the detected emission results from shock heated gas since the
\sii/\ha\ ratio exceeds the empirical SNR criterion value of 
0.4--0.5. Note that the \HII\ region found in the low ionization
images shows a \sii/\ha\ ratio of $\sim$0.2. The eastern filament lies
very close to this \HII\ region. The morphological differences between
the low and medium ionization lines provide evidence for significant
inhomogeneities and density variations in the ambient medium. Hester
et al. (1987) suggested that the presence of such inhomogeneities and
density variations would mainly affect the recombination zone where
the low ionization lines are produced and it could also explain the
\oiii/\ha\ ratio variations seen in the long--slit spectra.
  \begin{table}
        \caption[]{Relative line fluxes.}
         \label{sfluxes}
         \begin{flushleft}
         \begin{tabular}{lllllll}
     \hline
 \noalign{\smallskip}
                & Ia  & Ib \cr
\hline
Line (\AA) & F$^{\rm a,b}$ & F$^{\rm a,b}$ \cr
\hline
4861 \hbeta\ & 22 (3)$^{\rm c}$ & 15 (6) \cr
\hline
4959 [OIII]  & 70 (9) & 4 (2) \cr
\hline
5007 [OIII]  & 221 (28) & 23 (9) \cr
\hline  
6548 \nii\    & 39 (13)  & 21 (16) \cr
\hline
6563 \ha\     & 100 (30) & 100 (75) \cr
\hline
6584 \nii\    & 142 (44) & 75 (57) \cr
\hline
6716 \sii\    & 44 (15) & 41 (34) \cr
\hline
6731 \sii\    & 38 (13) & 30 (25) \cr
\hline
\hline
Absolute \ha\ flux$^{\rm d}$ & 7.9 & 19 \cr
\hline
\ha /\hbeta\   & 4.5 (3) & 6.7 (6) \cr
\hline
\sii/\ha\ 	& 0.82 (17) & 0.71 (36) \cr
\hline 
F(6716)/F(6731)	& 1.2 (10) & 1.4 (20) \cr
\hline 
\end{tabular}
\end{flushleft}
 ${\rm ^a}$ Uncorrected for interstellar extinction 

${\rm ^b}$ Listed fluxes are a signal to noise weighted
average of the individual fluxes

${\rm ^c}$ Numbers in parentheses represent the signal to noise ratio 
of the quoted fluxes

$^{\rm d}$ In units of \flux\\\
${\rm }$ All fluxes normalized to F(\ha)=100
\end{table}
\par
An interstellar extinction c at positions Ia and Ib (see
Table~\ref{sfluxes}), of 0.62 ($\pm$ 0.33) and 1.10 ($\pm$ 0.17) or an
A$_{\rm V}$~of 1.27 ($\pm$ 0.67) and 2.26 ($\pm$ 0.35) were measured,
respectively. We have also determined the electron density measuring the density sensitive line ratio of
\siirat. The densities we measure are below 400 \dens. Assuming that
the temperature is close to 10$^{4}$ K, it is possible to estimate
basic SNR parameters. The remnant under investigation is one of the
least studied remnants and thus, the current stage of its evolution is
unknown. The possibilities that the remnant is still in the
adiabatic phase or in the radiative phase of its
evolution cannot be excluded and will be examined in the following. The
preshock cloud density n$_{\rm c}$ can be measured by using the
relationship (Dopita 1979)

\begin{equation}
{\rm n_{[SII]} \simeq\ 45\ n_c V_{\rm s}^2}~{\rm cm^{-3}},
\end{equation}

where ${\rm n_{[SII]}}$ is the electron density derived from the
sulfur line ratio and V$_{\rm s}$ is the shock velocity into the
clouds in units of 100 \vel.  Furthermore, the blast wave energy can be
expressed in terms of the cloud parameters by using the equation given
by McKee \& Cowie (\cite{mck75})

\begin{equation}
{\rm E_{51}} = 2 \times 10^{-5} \beta^{-1} 
{\rm n_c}\ V_{\rm s}^2 \ 
{\rm r_{s}}^3 \ \ {\rm erg}.
\end{equation}

The factor $\beta$ is approximately equal to 1 at the blast wave
shock, ${\rm E_{51}}$ is the explosion energy in units of 10$^{51}$
erg and {\rm r$_{\rm s}$} the radius of the remnant in pc. By using
the upper limit on the electron density of 400 \dens, which was
derived from our spectra, we obtain from Eq. (1) that ${\rm n_c}
V_{\rm s}^2 < 8.9$. Then Eq. (2) becomes ${\rm E_{51}} < 0.08~{\rm
D_{1 kpc}^3}$, where ${\rm D_{1 kpc}}$~the distance to the remnant in
units of 1 kpc.
\par
An estimated value of N$_{\rm H} \sim 5 \times 10^{21}$~cm$^{-2}$
is given by Dickey \& Lockman (1990) for the 
column density in the direction of \gsnr. Considering
that (Ryter et al. 1975)

\begin{equation}
{\rm N_{H}} = 6.8~(\pm 1.6) \times 10^{21}~{\rm E_{B-V}}~{\rm cm}^{-2},
\end{equation}
 
where A$_{{\rm V}} = 3.1 \times {\rm E_{B-V}}$~(Kaler 1976), we obtain
an N$_{{\rm H}}$~of $3 \times 10^{21}~{\rm cm}^{-2}$~and $5 \times
10^{21}~{\rm cm}^{-2}$~for the two c values calculated from our
spectra, respectively. A column density of $2_{-2}^{+1}
\times 10^{21}~{\rm cm}^{-2}$~was measured towards an X--ray binary system,
located 1.2\degr\ east of
\gsnr\ at a distance of $\sim$3.1 kpc (Rib\'{o} et al. 1999). Since,
there are no other measurements of the interstellar density n$_{0}$,
values of 0.1 and 1.0 will be examined. Following the result of
Eq. (2) and assuming the typical value of 1 for the supernova
explosion energy (E$_{51}$), we find that the remnant may lie at
distances greater than 2.3 kpc. Then, the lower interstellar density
of $\sim$0.1 cm$^{-3}$~suggests that the column density is greater
than $7 \times 10^{20}~{\rm cm}^{-2}$, while for n$_{0} \approx 1~{\rm
cm}^{-3}$~it becomes greater than $7 \times 10^{21}~{\rm
cm}^{-2}$. Combining the previous results and assuming that the column
density is found in the range of $2 - 5 \times 10^{21}~{\rm cm}^{-2}$,
then the lower interstellar density seems to be more probable.
\par
Using the results of Cioffi et al. (1988) and the above range of
parameters for n$_{0}$~of 0.1 and 1 cm$^{-3}$, the pressure driven
snowplow (PDS) radii are calculated as 38 and 14 pc, respectively. The
radii, when compared with the estimated SNR radius of 17 ${\rm D_{2.3
kpc}}$~pc may suggest that the PDS phase has not begun yet. The
derived velocities of the main shock front at the beginning of this
stage are $\sim$300--400 km s$^{-1}$. Their comparison with the
estimated shock velocities into the interstellar clouds result in a
density contrast of $\sim$10 between the ambient medium and the
interstellar $``$clouds''.  However, since neither the distance nor
the interstellar medium density are accurately known, we cannot
confidently determine the current stage of evolution of \gsnr.
\section{Conclusions}
The faint supernova remnant \gsnr\ was observed for the first time in
major optical emission lines. The images show filamentary and diffuse
emission structures. The bright \oiii\ filament is very well
correlated with the remnant's radio emission at 1400 and 4850 MHz
suggesting their association. The flux calibrated images and the
long--slit spectra indicate that the emission arises from shock heated
gas. The observed optical filamentary structure provides some evidence
for significant inhomogeneities in the ambient medium, implying that
the main blast wave propagates into an inhomogeneous medium. 
\begin{acknowledgements}
\end{acknowledgements}
The authors would like to thank the referee for his comments and
suggestions which helped to clarify, and enhance, the scope of this
paper. Skinakas Observatory is a collaborative project of the
University of Crete, the Foundation for Research and Technology-Hellas
and the Max-Planck-Institut f\"ur Extraterrestrische Physik.  This
research has made use of data obtained through the High Energy
Astrophysics Science Archive Research Center Online Service, provided
by the NASA/Goddard Space Flight Center.
%

\end{document}